\documentclass[12pt]{iopart}

\usepackage{graphicx}
\usepackage{dcolumn}
\usepackage{bm}
\usepackage{xcolor}
\usepackage{amsmath}
\usepackage{amssymb}
\usepackage{textgreek}
\usepackage{appendix}
\usepackage{mathtools}

\begin{document}

\title[Phase-slip centers as cooling engines]{Novel results obtained by modeling of dynamic processes in superconductors: phase-slip centers as cooling engines}

\author{Iris Mowgood, Serafim Teknowijoyo, Sara Chahid and Armen Gulian}

\address{Advanced Physics Laboratory, Institute for Quantum Studies, Chapman University, Burtonsville, MD 20866 and Orange, CA 92866}
\ead{irismowgood@chapman.edu and gulian@chapman.edu}
\vspace{10pt}

\begin{abstract}
Based on a time-dependent Ginzburg-Landau system of equations and finite
element modeling, we present novel results related with the physics of
phase-slippage in superconducting wires surrounded by a non-superconductive
environment. These results are obtained within our previously reported
approach related to superconducting rings and superconductive gravitational
wave detector transducers. It is shown that the phase-slip centers (PSCs)
can be effective in originating not only positive but also negative thermal
fluxes. With an appropriate design utilizing thermal diodes, PSCs can serve as
cryocooling engines. Operating at $T\sim 1$ \textrm{K} cryostat
cold-finger, they can achieve sub-Kelvin temperatures without using $^3$He.

\end{abstract}

\noindent{\it Keywords\/}: TDGL equations, finite element modeling, phase-slip centers,
negative phonon fluxes, cryocooling, thermal diodes
\maketitle

\section{Introduction}

\begin{figure}
    \centering
    \includegraphics[width=\linewidth]{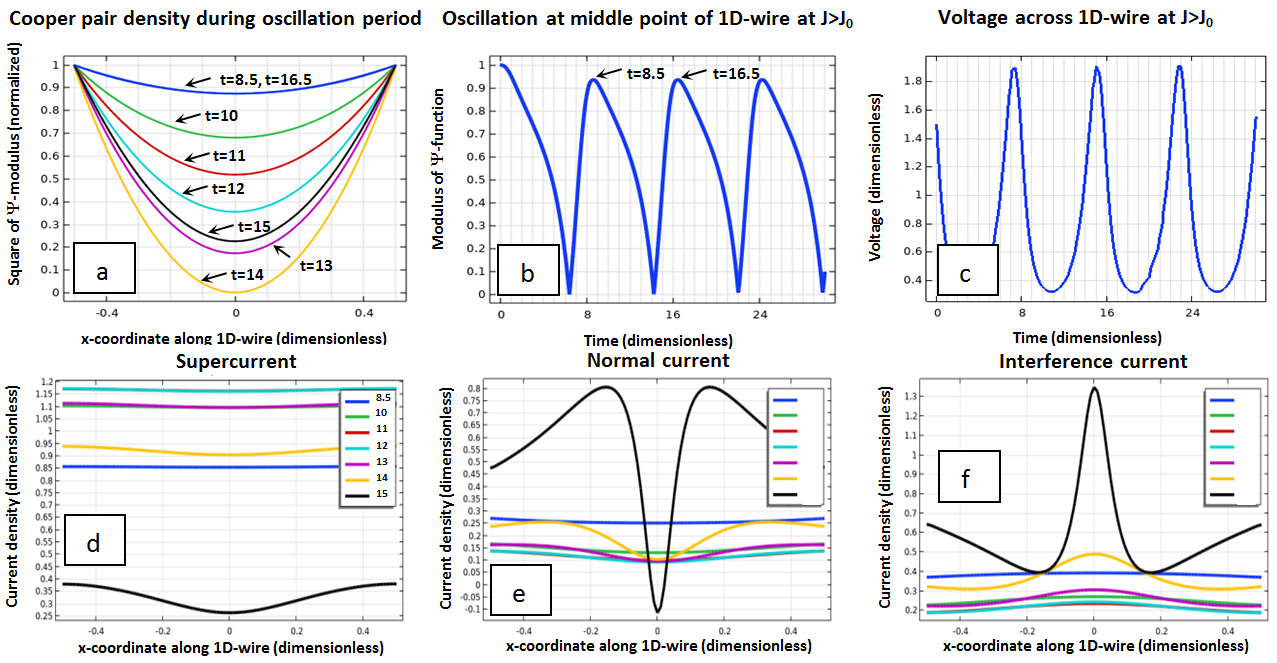}
    \caption{(\textbf{a}) Temporal oscillations of Cooper pair density $%
\left\vert \Psi (x,t)\right\vert ^{2}$ in the $1D-$wire at $J=1.5$ ($%
J_{0}\approx 1$). Moments of time marked by arrows correspond to those in
panel (\textbf{b}). (\textbf{b}) Time oscillations of $\left\vert \Psi
(x,t)\right\vert $ at the central point of PSC ($x=0$). (\textbf{c}) Voltage
between the ends of the $1D-$wire. (\textbf{d}) Supercurrent, (\textbf{e})
normal and (\textbf{f}) interference currents for various characteristic
moments of PSC oscillation (color coding is the same in panels \textbf{a}, 
\textbf{d}, \textbf{e} and \textbf{f}).}
    \label{fig1}
\end{figure}

Thin superconducting filaments (formally, $1D-$wires) enter into a unique,
so-called phase-slip state (which oscillates in time) when dc-biased above a
certain critical current: $J_{dc}>J_{0}$ (see, e.g., \cite{TidecksBook,BezryadinBook,Likharev70,GulianBook,KopninBook,GulianCOMSOL} and refs.
therein). These states have been explored for decades \cite{TidecksBook} and the
detailed understanding and description of microscopic mechanisms of
phase-slippage were achieved via analysis of the time-dependent
Ginzburg-Landau (TDGL) equations \cite{BezryadinBook,Likharev70,GulianBook,KopninBook,GulianCOMSOL,ScottBook}. These equations describe the
behavior of the wavefunction of Cooper pairs, as well as the normal
electrons (in conjunction with Maxwell's equations). Based on the most
complete form of the TDGL equations \cite{GulianCOMSOL} (which includes the
interference current \cite{Gulian87}) one can achieve a proper description of
physical processes in any BCS-type superconductor. As was established by the
research community \cite{Skocpol74}, the Cooper pair wavefunction phase
difference across the middle point of the superconducting wire jumps down by 
$2\pi $ periodically in time. At the moments of these jumps, the modulus of
the wavefunction in this middle point becomes zero. Since the square of
this wavefunction modulus is the density of Cooper pairs, supercurrent
varies accordingly, i.e., oscillates, and because of charge conservation in
the metal, all other current components (normal and interference) are also
oscillating, as well as the voltage between the ends of the wire. Figure \ref{fig1}
demonstrates these features.

As follows from Fig.~\ref{fig1} (\textbf{c}), the time-averaged voltage is non-zero,
which means that the filament at $J_{dc}>J_{0}$ is in a resistive state: $%
\bar{R}\neq 0$\ (hereafter, bars indicate time averaging). The overall
behavior of PSCs can be characterized as a quantum phenomenon similar to
traditional Josephson junctions (or weak links). For weakly coupled
superconductors, the  voltage $V_{dc}$ applied across the junction yields a
current oscillating with the Josephson frequency: $\omega _{J}\propto V_{dc}$%
. In the case of PSCs, we apply a dc current, and the appearing voltage
oscillates in time (see Fig.~1 (\textbf{c})) with frequency $\omega
_{\mathrm{PSC}}\propto V_{dc}$. In view of this analogy, PSCs can be qualified as 
``strong" weakly-coupled systems: there is
no actual weak link, and the central point plays the role of the
``weak point" because of symmetry. In this
article we demonstrate that phonon emission from PSCs is of non-trivial
character, and can be combined with thermal diodes to serve as a cryocooler
driving engine.

\section{Methods}

The finite element modeling based on the Mathematical Module of COMSOL
Multiphysics package proved to be a very productive tool for solving
the nonlinear time-dependent Ginzburg-Landau (TDGL) equations in various
studies of dynamic effects occuring in superconductors affected by
external fields \cite{Moshchalkov95,Baelus00,Baelus01,Baelus02,Chibotaru05,Vodolazov02,Vodolazov03,Peng14,Zha11,Vodolazov04}. In particular, our group analyzed the
behavior of $3D-$superconducting rings in varying magnetic fields and
discovered flux non-conservation in nanorings \cite{Mowgood22}%
; on the same basis, it was concluded that mutually interacting vertical
stacks of rings can serve as effective superconducting transducers for
gravitational wave detectors \cite{Gulian21}. In this report, we
focus on the $1D-$superconducting wires and explore dynamics of phase-slip
centers (PSCs) in them. The main topic of exploration is the phonon emission
by PSCs. For such a task, using the finite gap version of TDGL is mandatory.
We use the most complete set of these equations, which also includes the
interference current component \cite{GulianCOMSOL,Gulian87}.

The generalization of GL- equations for time-dependent problems took a 
considerable effort. The first successful step was by Schmid \cite{Schmid66}%
, who came to the conclusion that the proper equation for the $\Psi $%
-function is not like the Schr\"{o}dinger equation, but rather has a
structure similar to the diffusion equation. Later, this result was confirmed
by \'{E}liashberg and Gor'kov \cite{GorkovEliashberg1968} on the basis of Green's function
approach to superconductivity. Interestingly, at that point, a closed system
of TDGL equations resulted for gapless superconductors \cite{Abrikosov1961} only.
Later, after the development of a more powerful energy-integrated Green's
function technique for kinetics of non-equilibrium superconductors \cite{Schmid66,Eliashberg72,Larkin77,Gulian89}, 
the TDGL equations were derived for
finite gap superconductors $\cite{Gulian89,Golub76,Kramer78,Schon79,Hu80,Watts-Tobin81,Schmid81}$.

The first dynamic equation is for the order parameter $\Delta =\left\vert
\Delta \right\vert \exp (i\theta )$:

\begin{equation}
\begin{split}
-\frac{\pi }{8T_{c}}\frac{1}{\sqrt{1+(2\tau _{\varepsilon }\left\vert \Delta
\right\vert )^{2}}}\left( \frac{\partial }{\partial t}+2i\varphi +2\tau
_{\varepsilon }^{2}\frac{\partial \left\vert \Delta \right\vert ^{2}}{%
\partial t}\right) \Delta +\frac{\pi }{8T_{c}}\left[ D\left( \mathbf{\nabla }%
-2i\mathbf{A}\right) ^{2}\right] &\Delta \\
+\left[ \frac{T_{c}-T}{T_{c}}-\frac{%
7\zeta (3)\left\vert \Delta \right\vert ^{2}}{8(\pi T_{c})^{2}}+P(\left\vert
\Delta \right\vert )\right] &\Delta =0.  \label{1}    
\end{split}
\end{equation}%
Here the theoretical
units $\hbar =c=e=k_{B}=1$ are used. $\mathbf{A}$ and $\varphi $ are vector
and scalar potentials of the electromagnetic field, $\tau _{\varepsilon }$
is the electron-phonon relaxation time, $D$ is the electronic diffusion
coefficient, $\zeta (3)$ is the Riemann zeta function, and $P(\left\vert
\Delta \right\vert )$ is the non-equilibrium phonon term (in absence of
phonon feedback, $P(\left\vert \Delta \right\vert )\equiv 0$) which will be
specified below. The order parameter $\Delta $ ($\left\vert \Delta
\right\vert $ is the superconducting energy gap) is proportional to the
original Ginzburg-Landau $\Psi $-function (which can be normalized so that its
squared modulus is equal to the density of pair condensate). Equation (\ref%
{1}) describes the behavior of the Cooper-pair condensate taking into
account inelastic electron-phonon collisions. In the case of very strong
inelastic collisions ($\tau _{\varepsilon }\rightarrow 0$), Eq.(\ref{1})
converts into its gapless form, where the relaxation of $\Delta $ takes
place only due to the condensate itself demonstrating the self-restoring
property of usual Bose-condensates.

The second dynamic equation is for the electric current density, $\mathbf{j}$%
, which in the same approximation ($\tau _{\varepsilon }\neq 0$) should be
written as%
\begin{equation}
\begin{split}
\mathbf{j}&=\mathbf{j}_{s}+\mathbf{j}_{n}+\mathbf{j}_{int}=\frac{\pi \sigma
_{n}}{4T}\mathbf{Q}\left( \left\vert \Delta \right\vert ^{2}\mathbf{-}2\tau
_{\epsilon }\frac{\partial \left\vert \Delta \right\vert ^{2}}{\partial t}%
\right) \\
&+\sigma _{n}\mathbf{E}\left\{ 1+\frac{\left\vert \Delta \right\vert 
}{2T}\frac{\sqrt{1+(2\tau _{\epsilon }\left\vert \Delta \right\vert )^{2}}}{%
2\tau _{\epsilon }\left\vert \Delta \right\vert }\left[ K\left( \frac{2\tau
_{\epsilon }\left\vert \Delta \right\vert }{\sqrt{1+(2\tau _{\epsilon
}\left\vert \Delta \right\vert )^{2}}}\right) -E\left( \frac{2\tau
_{\epsilon }\left\vert \Delta \right\vert }{\sqrt{1+(2\tau _{\epsilon
}\left\vert \Delta \right\vert )^{2}}}\right) \right] \right\} ,  \label{2}    
\end{split}
\end{equation}%
where $\mathbf{E}\mathbf{=}\mathbf{-}\mathbf{\dot{A}}\mathbf{-}\mathbf{%
\nabla }\varphi $ and $\mathbf{Q}=\mathbf{-}2\mathbf{A}+\nabla \theta $. In
Eq.(\ref{2}), $K(x)$ and $E(x)$ are the complete elliptic integrals of the
first and second type, respectively. Since $[K(x)-E(x)]/x\rightarrow 0$ at $%
x\rightarrow 0$, one can recognize that for $\tau _{\varepsilon }\rightarrow 0
$, as in the case of Eq.~(\ref{1}), Eq.~(\ref{2}) yields the gapless limit.
This corresponds to the two-fluid model of superconductivity \cite{Gorter34}: $\mathbf{j}=\mathbf{j}_{s}+$ $\mathbf{j}_{n},$ where $%
\mathbf{j}=\mathbf{j}_{s}[\pi \sigma _{n}\left\vert \Delta \right\vert
^{2}/(4T)]\mathbf{Q}$ and $\mathbf{j}_{n}=\sigma _{n}\mathbf{E}$ ($\sigma
_{n}$ is the conductivity of normal excitations in the superconductor). The
``upgraded" version of the current in case of finite gap superconductors
corresponds to the interference between superfluid and normal motions of
electrons \cite{Gulian86}. With these interference terms included into the
expression for the current, Eq.~(\ref{2}) acquires the same level of
accuracy as Eq.~(\ref{1}) and the TDGL system can be used for a quantitative
description of effects in finite-gap superconductors.

Next, we consider the function $P(\left\vert \Delta \right\vert )$ in (\ref%
{1}). As shown in \cite{GulianBook,GulianCOMSOL}, this function has the form%
\begin{equation}
P(|\Delta |)=-2\tau _{\varepsilon }~\mathrm{Re}\int_{0}^{\infty }d\varepsilon {%
\frac{\Gamma (\varepsilon )}{\sqrt{(\varepsilon +i\gamma )^{2}-|\Delta |^{2}}%
}}\,,  \label{3}
\end{equation}%
where $\gamma =(2\tau _{\varepsilon })^{-1}$and $\Gamma (\varepsilon )$ is
related with the nonequilibrium population of phonons $\delta N_{\omega _{%
\mathbf{q}}}$ via 
\begin{equation}
\Gamma (\varepsilon )={\frac{\pi \lambda }{2(up_{F})^{2}}}\int_{0}^{\infty
}\omega _{\mathbf{q}}^{2}d\omega _{\mathbf{q}}\int_{|\Delta |}^{\infty
}d\varepsilon ^{\prime }\delta (\varepsilon ^{\prime }+\varepsilon -\omega _{%
\mathbf{q}})(u_{\varepsilon }u_{\varepsilon ^{\prime }}+v_{\varepsilon
}v_{\varepsilon ^{\prime }}){}\times (1-n_{\varepsilon }-n_{\varepsilon
^{\prime }})\delta N_{\omega _{\mathbf{q}}}.  \label{4}
\end{equation}%
In (\ref{4}), $\lambda $ is the dimensionless electron-phonon interaction
constant, $u$\ is the speed of sound in the superconductor, $p_{F}$\ is the
electrons Fermi momentum, $\omega _{\mathbf{q}}$ denotes phonon frequency
with the momentum $\mathbf{q}$, $u_{\varepsilon }=|\varepsilon |\theta
(\varepsilon ^{2}-|\Delta |^{2})/\sqrt{\varepsilon ^{2}-|\Delta |^{2}}$ is
the BCS density of states for electrons, $v_{\varepsilon }=u_{\varepsilon
}|\Delta |/\varepsilon ,$ and $n_{\varepsilon }$ is the distribution
function of nonequilibrium electrons. The nonequilibrium phonon distribution
function can be found from the kinetic equation%
\begin{equation}
{\frac{\partial }{\partial t}}(\delta N_{\omega _{\mathbf{q}}})=I(N_{\omega
_{\mathbf{q}}})+L(N_{\omega _{\mathbf{q}}}),  \label{5}
\end{equation}%
where $I(N_{\omega _{\mathbf{q}}})$ is the phonon-electron collision
integral, and $L(N_{\omega _{\mathbf{q}}})$ is the operator describing the
phonon exchange of a superconductor with its environment (the heat-bath). In
the simplest approximation \cite{Chang77,Chang86}, the latter
may be defined as 
\begin{equation}
L(N_{\omega _\mathbf{{q}}})\approx -{\frac{\delta N_{\omega _{\mathbf{q}}}}{\tau _{%
\mathrm{es}}}}\,,  \label{6}
\end{equation}%
where $\tau _{\mathrm{es}}\sim d/u$ is the phonon escape time (into the
heat-bath), and $d$ is the characteristic lateral dimension of the
superconductor. If $\tau _{\mathrm{es}}\rightarrow 0,$the phonons tend to
equilibrium, $\delta N_{\omega _{\mathbf{q}}}\rightarrow 0$. However, in
many practically important cases, $\tau _{\mathrm{es}}$\ is finite, and Eq.(%
\ref{5}) should be solved jointly with Eq.~(\ref{1}) and Eq.~(\ref{2}). In
the so-called ``generalized local equilibrium approximation", ${\partial }%
(\delta N_{\omega _{\mathbf{q}}})/\partial t=0$, and the solution for $%
\delta N_{\omega _{\mathbf{q}}}$ could be obtained from Eq.~(\ref{5}) \cite{GulianBook,GulianCOMSOL}. Leaving this case for future consideration, we will assume here
the fulfillment of the condition of free phonon exchange with the
environment, $\tau _{\mathrm{es}}\rightarrow 0$. Under this condition,
excess phonons are generated during the energy dissipation in
superconductors; however, they freely outflux without any dynamic feedback.
The phonon flux from the volume $\mathfrak{V}$ in a spectral range $d\omega
_{\mathbf{q}}$ is given by the expression \cite{AndersonBook} (see also \cite{GulianBook,GulianCOMSOL}%
): 
\begin{equation}
dW_{\omega _{\mathbf{q}}}=I(N_{\omega _{\mathbf{q}}}^{0})^{(ph-e)}\rho
(\omega _{\mathbf{q}})d\omega _{\mathbf{q}},  \label{6aa}
\end{equation}%
where $\rho (\omega _{\mathbf{q}})=\mathfrak{V}\omega _{\mathbf{q}%
}^{3}/(2\pi ^{2}u^{3})$ and the collision integral of phonons with electrons
is:%
\begin{equation}
\begin{split}
I(N_{\omega _{\mathbf{q}}}^{0}) 
&=\frac{\pi \lambda }{2}\frac{\omega _{D}}{%
\varepsilon _{F}}\int_{|\Delta |}^{\infty }d\varepsilon
\int_{|\Delta |}^{\infty }d\varepsilon ^{\prime }\\
&\times \Big\{\Phi
^{rec}(\varepsilon ,\varepsilon ^{\prime })[(N_{\omega _{\mathbf{q}%
}}^{0}+1)n_{\varepsilon }n_{\varepsilon ^{\prime }}-N_{\omega _{\mathbf{q}%
}}^{0}(1-n_{\varepsilon })(1-n_{\varepsilon ^{\prime }})]\delta (\varepsilon
+\varepsilon ^{\prime }-\omega _{\mathbf{q}})  \\
&+2\Phi ^{rel}(\varepsilon ,\varepsilon ^{\prime })[(N_{\omega _{\mathbf{q}%
}}^{0}+1)n_{\varepsilon }(1-n_{\varepsilon ^{\prime }})-N_{\omega _{\mathbf{q%
}}}^{0}(1-n_{\varepsilon })n_{\varepsilon ^{\prime }}]\delta (\varepsilon
-\varepsilon ^{\prime }-\omega _{\mathbf{q}})\Big\}  \label{7bb}    
\end{split}
\end{equation}%
Here $\omega _{D}$ the Debye frequency, $\varepsilon _{F}$\ the Fermi
energy, and $\Phi ^{rec}$ and $\Phi ^{rel}$\ are functions related to the
densities of states and coherence factors of electrons in superconductors: 
\begin{eqnarray}
\Phi ^{rec}(\varepsilon ,\varepsilon ^{\prime }) &=&\frac{\varepsilon }{%
\sqrt{\varepsilon ^{2}-|\Delta |^{2}}}\frac{\varepsilon ^{\prime }}{\sqrt{%
\varepsilon ^{\prime 2}-|\Delta |^{2}}}\left( 1+\frac{|\Delta |^{2}}{%
\varepsilon \varepsilon ^{\prime }}\right) ,  \label{8aa} \\
\ \Phi ^{rel}(\varepsilon ,\varepsilon ^{\prime }) &=&\frac{\varepsilon }{%
\sqrt{\varepsilon ^{2}-|\Delta |^{2}}}\frac{\varepsilon ^{\prime }}{\sqrt{%
\varepsilon ^{\prime 2}-|\Delta |^{2}}}\left( 1-\frac{|\Delta |^{2}}{%
\varepsilon \varepsilon ^{\prime }}\right) .  \label{8bb}
\end{eqnarray}%
Most importantly, in (\ref{7bb}), $n_{\varepsilon \text{ }}$is the \textit{%
nonequilibrium} distribution function of electron excitations, while $%
N_{\omega _{\mathbf{q}}}^{0}$\ is the \textit{equilibrium} phonon
distribution function. In writing (\ref{7bb}) it is assumed that $%
n_{\varepsilon \text{ }}=n_{-\varepsilon \text{ }},$ i.e., electron ($%
\varepsilon >0$) and hole ($\varepsilon <0$) population branch imbalance is absent. In
the context of the problem under consideration, this restriction is not
principal. In thermodynamic equilibrium at temperature $T$, $n_{\varepsilon
}=n_{\varepsilon }^{0}=1/[\exp (|\varepsilon |/T)+1]$, while $N_{\omega _{%
\mathbf{q}}}^{0}=1/[\exp (\omega _{\mathbf{q}}/T)-1]$. Substitution of these
functions into $I(N_{\omega _{\mathbf{q}}}^{0})$ yields identical zero, which
means that in equilibrium, the processes of phonon emission by electron
excitations (terms $\propto $ $(N_{\omega _{\mathbf{q}}}^{0}+1)$ in (\ref%
{7bb})) are exactly cancelled by the reciprocal processes of phonon
absorption by electrons (terms $\propto $ $N_{\omega _{\mathbf{q}}}^{0}$).
These processes of emission and absorption take place continuously in
superconductors at $T\neq 0$ because of thermodynamic fluctuations. Their
exact cancellation reflects the so-called principle of detailed balance.
This principle is violated when an external source drives the system
away from equilibrium (for example, by breaking Cooper pairs). Then $\delta
n_{\varepsilon }\equiv n_{\varepsilon }-n_{\varepsilon }^{0}>0$.
Substitution of $\delta n_{\varepsilon }>0$ into $I(N_{\omega _{\mathbf{q}%
}}^{0})$\ generates excess phonons, i.e., indicates the phonon outflow. If $%
\delta n_{\varepsilon }<0,$ then in some energy range the phonon flux is
negative, i.e., the ``phonon deficit
effect" \cite{Gulian80} takes place.

This brings us close to the most interesting case, where the behavior of $%
I(N_{\omega _{\mathbf{q}}}^{0})$\ at a PSC is determined by the sign-varying 
$\delta n_{\varepsilon }(t)$ function. This function is related with the
dynamics of $\left\vert \Delta (x,t)\right\vert $ via 
\begin{equation}
\delta n_{\varepsilon }(t)=\alpha \frac{\partial \left\vert \Delta
(x,t)\right\vert }{\partial t},  \label{9aa}
\end{equation}%
as follows from the derivation of the TDGL equations (see, e.g., \cite{GulianCOMSOL}%
). The proportionality coefficient $\alpha $\ in (\ref{9aa}) is positively
defined: $\alpha =(1/2T)\cosh ^{-2}(\varepsilon /2T)(R_{2}/N_{1})$, where $%
N_{1}(\varepsilon )=\mathrm{Re}\left\{ [\varepsilon +i\gamma ]/\sqrt{%
(\varepsilon +i\gamma )^{2}-\left\vert \Delta \right\vert ^{2}}\right\} ,$ $%
R_{2}(\varepsilon )=\mathrm{Re}\left[ \left\vert \Delta \right\vert /\sqrt{%
(\varepsilon +i\gamma )^{2}-\left\vert \Delta \right\vert ^{2}}\right] $. In
view of excellent correspondence between the theoretical description of PSCs
by TDGL and detected experimental data as of now, we can rely on the
applicability of Eq.~(\ref{9aa}) to other PSC-related phenomena, such as the
accompanying phonon emission which will be considered next. The
computational COMSOL code and the dimensionless form of the TDGL equations
used for obtaining the results below are described in detail in Appendices A
and B.

\section{Results and Discussion}

From the consideration of panel (\textbf{b}) in Fig.~1 and the discussion
above, one can draw a qualitative conclusion that the phonon fluxes related
to PSCs dynamically reverse their sign: a positive flux (generation of
phonons) is followed by a negative flux (absorption of phonons), and so on.
Figure 2 illustrates this.

\begin{figure}
    \centering
    \includegraphics[width=\linewidth]{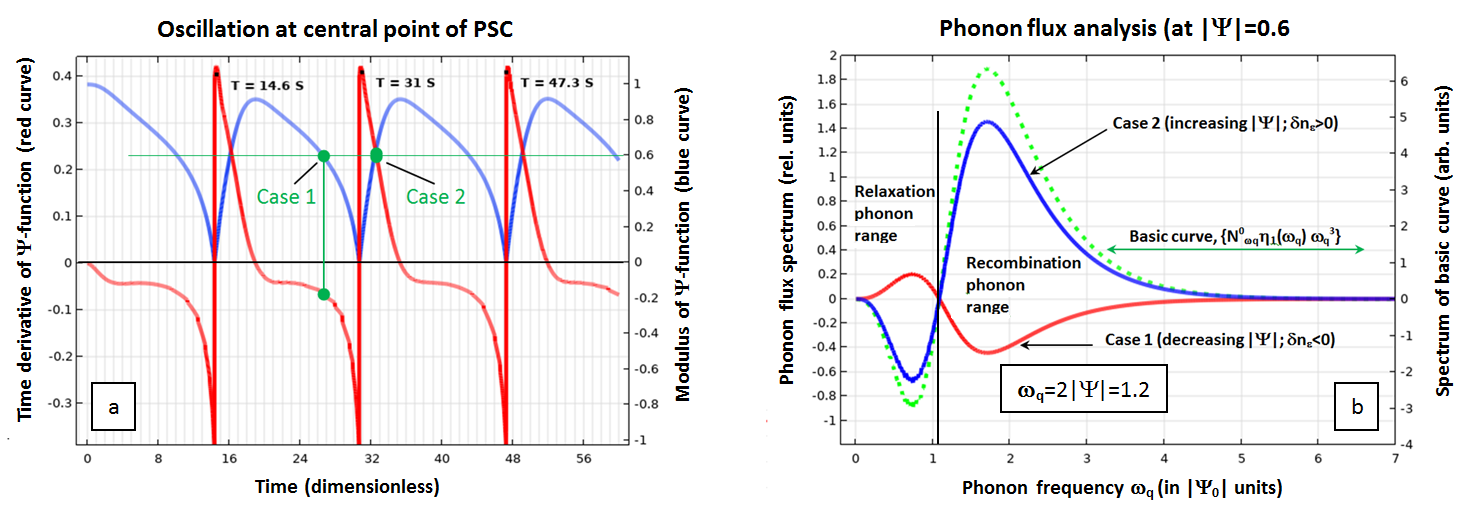}
    \caption{(\textbf{a}) This form of $\partial \left\vert \Psi (x,t)\right\vert
/\partial t$\ at a PSC (red curve) causes a pulsating phonon flux spectrum.
The flux reverses its sign during the period of PSC oscillation, Cases 1 and
2. (\textbf{b}) Curves corresponding to Eq.~(3). At a given moment of time,
independently of $\partial \left\vert \Psi \right\vert /\partial t$, the
spectrum of phonon flux has positive and negative segments, separated by the
frequency $\omega _{q}=2\left\vert \Psi \right\vert $. Additionally, it
depends on variables $\left\vert \Psi \right\vert $, $T$, and $\gamma $. The
curves shown are for values of $T=0.9T_{c}$\ and $\gamma =0.3$.}
    \label{fig2}
\end{figure}

The pulsation of reciprocating phonon fluxes has been predicted in Ref. \cite{Gulian86}, 
albeit at that time, no cooling mechanism was anticipated. Meanwhile,
the cooling engine is one step away from the theoretical findings plotted in
Fig.~2(\textbf{b}): we will have a prerequisite for a cooling engine in
which the PSC absorbs phonons from a certain part of its environment (the
medium to be cooled) and emits the phonons to another part of its
environment (the heat sink). Thermal diodes can serve perfectly for the 
separation of positive and negative fluxes required for this step, as will
be shown below.

\subsection{PSCs as cryocooling elements}

During the conditions of free exchange of phonons between the superconductor
and surrounding media, the energy spectrum of the phonon emission at
frequency $\omega _{\mathbf{q}}$ in the range $d\omega _{\mathbf{q}}$\ has
the form (\ref{6aa}), which can be represented as:%
\begin{equation}
dW_{\omega _{\mathbf{q}}}=\frac{\lambda \tau _{\varepsilon }\mathfrak{V}}{%
2\pi u^{3}T}\frac{\omega _{D}}{\varepsilon _{F}}\frac{\partial \left\vert
\Psi \right\vert }{\partial t}\left[ N_{\omega _{\mathbf{q}}}^{0}\eta
_{1}(\omega _{\mathbf{q}})\omega _{\mathbf{q}}^{3}\right] d_{\omega_{\mathbf{q}}}
\label{6cc}
\end{equation}%
Here $u$ is the acoustic phonon propagation speed, and the function $\eta
_{1}(\omega _{\mathbf{q}})$ is described by the formula%
\begin{equation}
\begin{split}
\eta _{1}(\omega _{\mathbf{q}})&=\frac{1}{4}\int_{0}^{\infty
}d\varepsilon \frac{P(\varepsilon )R_{2}(\varepsilon )}{N_{1}(\varepsilon
)\cosh ^{2}(\varepsilon /2T)}\\
&-\frac{1}{4}\int_{0}^{\infty
}d\varepsilon Q(\varepsilon )\left\{ \frac{R_{2}(\varepsilon \text{ }+\text{ 
}\omega _{\mathbf{q}})}{N_{1}(\varepsilon \text{ }+\text{ }\omega _{\mathbf{q%
}})\cosh ^{2}[(\varepsilon \text{ }+\text{ }\omega _{\mathbf{q}})/2T]}-\frac{%
R_{2}(\varepsilon )}{N_{1}(\varepsilon )\cosh ^{2}[\varepsilon /2T]}\right\}
,  \label{6ccc}    
\end{split}
\end{equation}%
where $P(\varepsilon )=N_{1}(\varepsilon )N_{1}(\omega _{\mathbf{q}%
}-\varepsilon )+R_{2}(\varepsilon )R_{2}(\omega _{\mathbf{q}}-\varepsilon ),$
and $Q(\varepsilon )=N_{1}(\varepsilon )N_{1}(\omega _{\mathbf{q}%
}+\varepsilon )-R_{2}(\varepsilon )R_{2}(\omega _{\mathbf{q}}+\varepsilon )$. 
The function $\eta _{1}(\omega _{\mathbf{q}})$ determines non-trivial
spectral properties of the phonon emission. The combination of this function
with other quantities determining the $\omega _{\mathbf{q}}$--dependence of
phonon energy fluxes is shown in Fig.~2(b). In accordance with it, at any
moment of time, the energy spectrum of the PSC phonon flux has negative and
positive segments. When $\partial \left\vert \Psi \right\vert /\partial t>0$
, positive segments correspond to phonon emission and negative segments
correspond to phonon absorption. When $\partial \left\vert \Psi \right\vert
/\partial t<0$, the situation is reversed. Integration over the energy
spectrum yields total energy release at a given moment of time. To find the
net energy release, one should also integrate over the period of
oscillation. As was mentioned in \cite{Gulian86}, and as can be deduced from the
plots in Fig.~2, the resultant value of the net energy release (which should
be close to the Joule heating by the dissipative components of the current
in the filament) is much smaller than the sole contributions of negative and
positive fluxes: they are mainly cancelling each other. Meanwhile,
separation of actions of negative and positive phonon fluxes via thermal
diodes can be used for designing a cryocooler.

\subsection{Role of thermal diodes}

For this task, one must analyze a little further how the phonons propagate
from the filament into the substrate and vice versa. Let us consider a
geometry where the superconducting filament is located between two plates,
as shown in Fig.~3. In this arrangement, the materials of the top and the
bottom plate are different from each other. In particular, the top plate has
acoustic density $\rho u$ lower than that of the superconductor, while the
bottom plate (substrate) has acoustic density $\rho u$ higher than the superconductor.

\begin{figure}
    \centering
    \includegraphics[width=.5\linewidth]{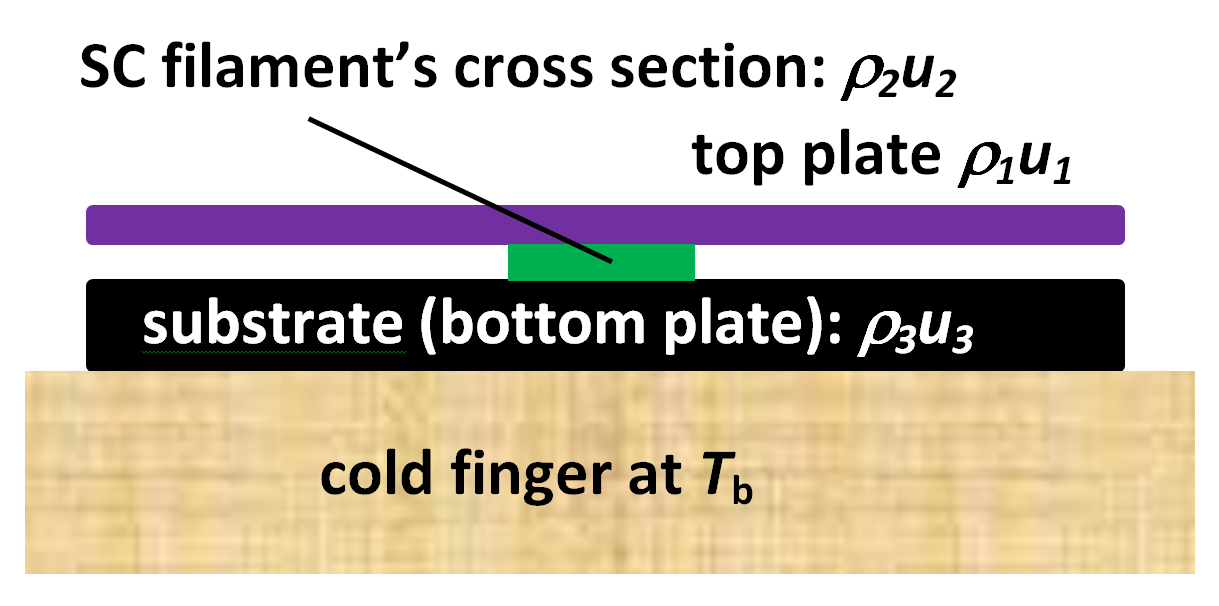}
    \caption{Cross-sectional view of the cooler design based on PSC-filament with
two surrounding plates.  Acoustic densities should satisfy the relation: $%
\rho _{1}u_{1}<\rho _{2}u_{2}<\rho _{3}u_{3}$.}
    \label{fig3}
\end{figure}

Here $\rho $ is the mechanical density of the material, and $u$ is the speed
of acoustic longitudinal phonons in it. The difference between acoustical
densities will yield the effect of Kapitza resistance between layers in Fig.
3. This interfacial Kapitza resistance to heat flow \cite{TidecksBook,AndersonBook} reflects
the acoustical mismatch at the interfaces and depends on the direction of
the heat transfer. The parameter $\rho u$ in acoustics plays the same role
as the refraction index $n$ in  optics. Phonons propagating from the
acoustically higher density material into an acoustically lower density
material suffer complete internal reflection at the boundary if their angle
of incidence exceeds a value determined by the ratio of acoustical
densities, thus contributing to the total Kapitza resistance. This
reflection is absent for the propagation of phonons in the opposite
direction, \textit{i.e.} from the acoustically lower density to higher density
material. Thus, the Kapitza resistance plays the role of a simple thermal
diode \footnote{
There are many other thermal diode designs described in the literature (see 
\cite{Melkonyan03} and references therein) but we are adopting the simplest
implementation here.} in the design shown in Fig.~3. Recall the presence
of two mechanisms during the action of the PSC in the filament: the emission
and the absorption of phonons. The cooling of the top plate is accomplished
by the free flow of thermal phonon fluxes into the filament to
preferentially fill-in the phonon deficit in the filament during the portion
of the cycle when this deficit exists while the Kapitza resistance prohibits
significant phonon influx into the top plate from the filament during the
portion of the cycle when excess phonons are generated in the filament.
These excess phonons, when generated in the filament by a PSC, freely enter
the substrate which is held at cold-finger temperature. Note that the
propagation of phonons from the substrate to the filament is suppressed by
the Kapitza resistance, so the substrate phonons are not allowed to
effectively fill-in the phonon deficit in the filament when it occurs. Thus
the filament with the actively oscillating PSC should serve as an effective
phonon pump from the top plate into the cryostat cold-finger.

\subsection{Cooling efficiency}

The phonon pump itself is not yet sufficient for cryocooler performance. One
should make sure that parasitic heat links do not impede its action. In our
case (Fig.~3), such a heat link is provided by the PSC filament itself.
If the top plate is at temperature $T_{\mathrm{cold}}$, and the bottom plate is at the
heat-bath temperature $T_{b}$, then the power of heat leakage through the
superconducting PSC filament is

\begin{equation}
P_\text{{leak}}=k\frac{S}{d}(T_{b}-T_{\text{cold}})  \label{11aa}
\end{equation}%
where $k$ is the thermal conductivity of the PSC filament material, $S$\ is
the surface of the area of the filament in thermal contact between top and
bottom plates in Fig.~3, and $d$\ is the thickness of the filament. This
value should be compared to the cooling power $P_{c}$ of the phonon pump.
Obviously, it cannot exceed the range of characteristic electric power
required for the existence of a single PSC\footnote{%
Application of higher electric power through the filament will lead to the
appearance of multiple PSCs, which is associated with more complex behavior
and is out of current consideration.}: $P_{c}%
\sim \varkappa P_{\mathrm{PSC}}=J_{dc}\times \bar{V}$, where $\varkappa<1$. 
The proper value of $\varkappa $\ can be computed
using the data obtained from our modeling by integration of phonon fluxes over the
frequency $\omega _{q}$ and over the oscillation period, provided the thermal diode
efficiency $p$ $\footnote{%
Each of the two interfaces will have their own $p$-factors, and we assume that
the one we introduced here is a certain average of them.}$ is known. This
quantity is also possible to compute via Kapitza boundary modeling. For
net cooling to take place, one must fulfill the condition%
\begin{equation}
\varkappa >k\frac{S}{dJ_{dc}\times \bar{V}}(T_{b}-T_{\text{cold}}).
\label{12aa}
\end{equation}%
In addition to this heat leakage, there will be another leakage during
the delivery of the electric power from the cryostat cold stage to the PSC
filament. However, if the PSC filaments are nearly identical, one can connect
them in series, so that the resistance of the group is high, and biasing
current is small enough to keep (super)conducting wires from the cold finger
thin (and accordingly, to keep the PSCs heatload small) while the cooling
power will be much larger for the whole group. Cooling from $T_{b}=1$ 
\textrm{K} to sub-K temperatures should be possible, if the value of $%
\varkappa $\ (\ref{12aa}) is large enough. This will be the focus of future
studies.

\section{Conclusion}

Thermal behavior of $1D-$superconducting wire in resistive states may be
suitable for cryocooler application. Unlike resistive states in normal
wires, where the electromagnetic energy is being converted into phonon
fluxes and absorbed by the heat sink, in superconducting wires with PSCs it
has peculiar dependence with positive and negative fluxes in certain
spectral regions. Moreover, these fluxes change their signs during the
oscillational period. The total phonon flux integrated over the period of
PSC is positive and is related with the resistive energy dissipation.
Spatial separation of these positive and negative phonon fluxes is possible via
thermal diodes. In this case positive phonon flux is being directed into the
heat sink, while the negative flux is being compensated by the extraction of
phonons from the medium to be cooled down. Total energy balance is still
positive and corresponds to electromagnetic energy dissipation; however,
we have here an analogue to situations where heating causes cooling \cite{Cleuren12,Mari12}.

\section*{Acknowledgments}
This work was supported by the ONR Grants N00014-19-1-2265 and N00014-21-1-2879.





\appendix

\section{Dimensionless equations}

For 1D-filaments, the general form of TDGL equations can be simplified,
since we have freedom of using 2 additional constraints and the freedom of
using arbitrary gauge for the electromagnetic potentials. Thus, one can
drop the vector potential, choosing all three components $%
A_{x}=A_{y}=A_{z}=0$ in Eqs. (\ref{1}) and (\ref{2}). By this choice, the
influence of magnetic induction $\mathbf{B}$ on the physical picture in
1D-filament is dropped. In fact, this physical picture is defined by the
value of total current through the filament, $J_{0}$. Since there is no
dependence of the current on the transverse coordinates, $j$ can depend only
on the coordinate $x$ along the filament. However, the condition $\mathrm{div}%
~\mathbf{j}=0$ in $1D$-cases implies $j=const$. This constant will serve as
an external variable of the problem. In the dimensionless form, Eq.~(\ref{1}%
) can be represented as

\begin{equation}
\nu \left( \frac{\partial }{\partial \tau }+i\text{\ }\varphi +\frac{1}{2}%
\delta ^{2}\frac{\partial \left\vert \Psi \right\vert ^{2}}{\partial \tau }%
\right) \Psi =\Psi ^{\prime \prime }+\left( 1-\left\vert \Psi \right\vert
^{2}+p\right) \Psi ,  \label{A1}
\end{equation}%
where the dimensionless order parameter is $\Psi =\Delta /\Delta _{0}$, $%
\Delta _{0}=\{8\pi ^{2}T_{c}^{2}\eta /[7\zeta (3)]\}^{1/2}$, ($\zeta
(3)\approx 1.2$\ is the Riemann $\zeta -$ function), $\nu =u/\sqrt{1+\delta
^{2}\left\vert \Psi \right\vert ^{2}}$, $\eta =(T_{c}-T)/T_{c}$, $\delta
=2\tau _{\varepsilon }\Delta _{0}$, $u=\pi ^{4}/[14\zeta (3)]\approx 5.798$, 
$\tau =tuD/\xi ^{2}\equiv t/t_{0}$, $t_{0}=\xi ^{2}/(Du)=\pi /[8u(T_{c}-T)]$, 
$\bar{\varphi}=2\varphi \xi ^{2}/(uD)\equiv 2\varphi t_{0}$, $\xi \equiv
\xi (T)=\{\pi D/[8(T_{c}-T)]\}^{1/2}$ ($\xi (T)$ is the ``dirty metal"
superconducting coherence length), $\bar{x}=x\mathbf{/}\xi (T)$, and $%
p=P(|\Delta |)/\eta $. In writing Eq.~\ref{A1}, we dropped the ``bar" symbols
(like $\bar{x}$) in the dimensionless quantities. Similarly, Eq.~(\ref{2})
acquires the form: 
\begin{equation}
\begin{split}
j=&-\left[ \frac{i}{2\left\vert \Psi \right\vert ^{2}}\left( \Psi ^{\ast
}\Psi ^{\prime }-\Psi \Psi ^{\ast \prime }\right) \right] \left( \left\vert
\Psi \right\vert ^{2}\mathbf{-}2\delta \sqrt{\frac{\eta }{u}}\frac{\partial
\left\vert \Psi \right\vert ^{2}}{\partial \tau }\right) \\
&-\varphi ^{\prime
}\left\{ 1+\frac{2}{\pi }\sqrt{\eta u}\left\vert \Psi \right\vert \frac{%
\sqrt{(\left\vert \Psi \right\vert \delta )^{2}+1}}{\left\vert \Psi
\right\vert \delta }\left[ K\left( \frac{\left\vert \Psi \right\vert \delta 
}{\sqrt{\left\vert \Psi \right\vert ^{2}\delta ^{2}+1}}\right) -E\left( 
\frac{\left\vert \Psi \right\vert \delta }{\sqrt{\left\vert \Psi \right\vert
^{2}\delta ^{2}+1}}\right) \right] \right\} .  \label{A2}   
\end{split}
\end{equation}%
with $\mathbf{\ }\bar{j}=j/j_{0}$, where $j_{0}=4u\sigma _{n}\left(
T_{c}-T\right) /\left[ \xi (T)\pi \right] $. The difference of elliptic
functions, as in \cite{Mowgood22}, with high-enough accuracy can be
replaced by the expression: $K(x)-E(x)\approx \left[ \ln (1+x)-\ln (1-x)%
\right] /2+(1-x)\ln (1-x)\equiv \ln \sqrt{(1-x^{2})}-x\ln (1-x)$.

\begin{equation}
\varphi ^{\prime }=\frac{-\left[ \frac{i}{2\left\vert \Psi \right\vert ^{2}}%
\left( \Psi ^{\ast }\Psi ^{\prime }-\Psi \Psi ^{\ast \prime }\right) \right]
\left( \left\vert \Psi \right\vert ^{2}\mathbf{-}2\delta \sqrt{\frac{\eta }{u%
}}\frac{\partial \left\vert \Psi \right\vert ^{2}}{\partial \tau }\right) -j%
}{\left\{ 1+\frac{2}{\pi }\sqrt{\eta u}\left\vert \Psi \right\vert \left[ -%
\frac{\sqrt{(\left\vert \Psi \right\vert \delta )^{2}+1}}{2\left\vert \Psi
\right\vert \delta }\ln \left( \left\vert \Psi \right\vert ^{2}\delta
^{2}+1\right) -\ln \left( 1-\frac{\left\vert \Psi \right\vert \delta }{\sqrt{%
\left\vert \Psi \right\vert ^{2}\delta ^{2}+1}}\right) \right] \right\} }.
\label{A3}
\end{equation}

One can notice that in case of $\tau _{\varepsilon }\rightarrow 0$
(in which case also $p=0$), Eqs. \ref{A1} and \ref{A2} reduce to the
well-known gapless case:%
\begin{equation}
u\text{ }\dot{\Psi}+i\text{\ }u\varphi \Psi =\Psi ^{\prime \prime }+\left(
1-\left\vert \Psi \right\vert ^{2}\right) \Psi ,  \label{A4}
\end{equation}%
\begin{equation}
\varphi ^{\prime }=\mathrm{Im}\left( \Psi ^{\ast }\Psi ^{\prime }\right) -j,
\label{A5}
\end{equation}%
as they should. Thus, in the 1D-case, we have 2 partial differential equations
for the real and the imaginary parts of the $\Psi -$function, and one
ordinary differential equation for the potential $\varphi $. Equation (\ref%
{A5}) can be written in integral form: 
\begin{equation}
\varphi (x,t)=\int_{-L_{0}/2}^{x}\left[ \mathrm{Im}\left( \Psi ^{\ast }\Psi
^{\prime }\right) -j\right] dx-\frac{1}{2}\int_{-L_{0}/2}^{L_{0}/2}\left[ 
\mathrm{Im}\left( \Psi ^{\ast }\Psi ^{\prime }\right) -j\right] dx  \label{A6}
\end{equation}%
where the constant of integration is chosen in such a way that the function
is antisymmetric on the $x$-axis at any moment of time. Equation (\ref{A3})
can be represented similarly. We stress that the potential $\varphi $\ (\ref%
{A6}) cannot be gauge-transformed by an arbitrary constant since the gauge
has been already declared above, and $\varphi $\ is, in fact, a
gauge-invariant quantity. The difference of $\varphi $\ at the ends of the
wire determines the voltage $V(t)=\varphi (L_{0}/2,t)-\varphi (-L_{0}/2,t)$
in the resistive state. Importantly, if the filament is attached to massive
superconducting banks, the phase $\theta $\ of the wavefunction $\Psi $\ at
the ends of the filament ($x=\pm L_{0}/2$) is related with potential $%
\varphi $\ by the relation:%
\begin{equation}
\dot{\theta}=-\varphi .  \label{A7}
\end{equation}%
Integrating (\ref{A7}) at $x=\pm L_{0}/2$ with the initial condition $\theta
(t=0)=0$, one can formulate boundary conditions for the $\Psi $-function: 
\begin{equation}
\mathrm{Re}\Psi |_{x=\pm L_{0}/2}=\cos \theta (\pm L_{0}/2),  \label{A8}
\end{equation}%
\begin{equation}
\mathrm{Im}\Psi |_{x=\pm L_{0}/2}=\sin \theta (\pm L_{0}/2).  \label{A9}
\end{equation}

As soon as the $\Psi -$function is determined, further evaluation of phonon
fluxes (\ref{6cc}) is straightforward.

\section{COMSOL coding}

For modeling, we used the Mathematics module of COMSOL 6.0. The time dependent 
Study 1 contains two equations. The first one is the general form PDE 
\begin{equation}
e_{a}\frac{\partial ^{2}\mathbf{u}}{\partial t^{2}}+d_{a}\frac{\partial 
\mathbf{u}}{\partial t}+\mathbf{\nabla }\cdot \mathbf{\Gamma }=\mathbf{f,}
\label{B1}
\end{equation}%
which is used for (\ref{B1}). Here $\mathbf{u}=[u1,u2]^{T}$, $u1=\mathrm{Re}%
\Psi $, $u2=\mathrm{Im}\Psi $,%
\begin{equation}
\mathbf{\Gamma }=\left( 
\begin{array}{c}
-u1x \\ 
-u2x%
\end{array}%
\right) ,\text{ \ \ }\mathbf{\nabla }=\left( \frac{\partial }{\partial x},%
\frac{\partial }{\partial y}\right) ,  \label{B2}
\end{equation}%

\begin{equation}
\mathbf{f}=\left( 
\begin{array}{c}
u1\ast (1-u1\symbol{94}2-u2\symbol{94}2)+\varphi \ast u2\ast \nu  \\ 
u2\ast (1-u1\symbol{94}2-u2\symbol{94}2)-\varphi \ast u1\ast \nu 
\end{array}%
\right) ,  \label{B3}
\end{equation}%

\begin{equation}
e_{a}=\left( 
\begin{array}{cc}
0 & 0 \\ 
0 & 0%
\end{array}%
\right) ,\text{ \ \ \ }d_{a}=\left( 
\begin{array}{cc}
(1+\delta \symbol{94}2\ast u1\symbol{94}2)\ast \nu  & \delta \symbol{94}%
2\ast u1\ast u2\ast \nu  \\ 
\delta \symbol{94}2\ast u1\ast u2\ast \nu  & (1+\delta \symbol{94}2\ast u2%
\symbol{94}2)\ast \nu 
\end{array}%
\right) .  \label{B4}
\end{equation}%
The scalar potential $\varphi $ which enters (\ref{B1}) is defined via set
of variables (\ref{B5})-(\ref{B10}):

\begin{equation}
\begin{split}
\varphi  =&intop1(\theta \ast ((u1\ast u2x-u2\ast u1x)\ast \\
&(1-coeff1\ast
(u1\ast u1t+u2\ast u2t))-J0)/(1+coeff2\ast Z(z)))-  \\
&-0.5\ast intop1(((u1\ast u2x-u2\ast u1x)\ast \\
&(1-coeff1\ast (u1\ast
u1t+u2\ast u2t))-J0)/(1+coeff2\ast Z(z))),  \label{B5}
\end{split}
\end{equation}%

\begin{equation}
coeff2=\frac{2}{\pi }\ast sqrt(\eta \ast u\ast (u1\symbol{94}2+u2\symbol{94}%
2))  \label{B6}
\end{equation}%
\begin{equation}
z=\delta \ast \frac{sqrt(u1\symbol{94}2+u2\symbol{94}2)}{sqrt(1+\delta 
\symbol{94}2\ast (u1\symbol{94}2+u2\symbol{94}2))}  \label{B7}
\end{equation}%

\begin{equation}
\theta =(dest(x)-x)>0  \label{B8}
\end{equation}%

\begin{equation}
coeff1=\frac{4}{u1\symbol{94}2+u2\symbol{94}2}\ast \delta \ast sqrt(\frac{%
\eta }{u})  \label{B9}
\end{equation}%
\begin{equation}
\nu =\frac{u}{sqrt(1+\delta \symbol{94}2\ast (u1\symbol{94}2+u2\symbol{94}2))%
}  \label{B10}
\end{equation}

Function $Z(x)$ in (\ref{B5}) is defined as:

\begin{equation}
Z(x)=\frac{1}{2\ast x}\ast \ln (1-x\symbol{94}2)-\ln (1-x)  \label{B11}
\end{equation}%
Initial conditions are: $u1=1$ and $u2=0$. Dirichlet boundary conditions are 
$u1=\cos (u3)$ and $u2=\sin (u3)$, \ where $u3$ is the phase of the wave
function $\Psi $. The values of phase are required only at the ends of the
wire; that's why it is enough to use Boundary ODEs and DAEs (bode), of which
we chose Distributed ODE 
\begin{equation}
e_{a}\frac{\partial ^{2}u3}{\partial t^{2}}+d_{a}\frac{\partial u3}{\partial
t}=f  \label{B12}
\end{equation}%
where $f=-\varphi $, $e_{a}=0$, $d_{a}=1$, with the initial values $u3=0$ at
the ends of the wire. This is the second equation of Study 1. These
equations enable one to calculate the behavior of $\Psi $-function, as well as
the gauge-invariant scalar potential $\varphi $. For plotting the results,
one needs $\nabla \varphi $, which, for enough accuracy, requires special
computation performed in the time dependent Study 2. It consists of an
algebraic equation $f(u4)=0$, with $f=4-withsol(^{\prime }sol1^{\prime
},fi,setval(t,t))$, which introduces the variable $u4$ via a distributed
ODE1:%
\begin{equation}
e_{a}\frac{\partial ^{2}u4}{\partial t^{2}}+d_{a}\frac{\partial u4}{\partial
t}=f  \label{B13}
\end{equation}%
where $f=-1$, $e_{a}=0$, $d_{a}=1$, with the initial values $u4=0$. This
allows us to compute current components shown in Fig.~1 with high enough
accuracy. Still, at plotting the curves it is advantegeous to use $%
centroid(-u4x)$ option for $\nabla \varphi $ to avoid boundary
peculiarities. At computations we use Extremely fine mesh with total degrees
of freedom less than $1000$.

\section*{References}
\bibliographystyle{iopart-num}

\providecommand{\noopsort}[1]{}\providecommand{\singleletter}[1]{#1}%
\providecommand{\newblock}{}

\end{document}